# Resonant quenching of Raman scattering due to out-of-plane $A_{1g}/A'_1$ modes in few-layer MoTe$_2$

K. Gołasa,[1,a)] M. Grzeszczyk,[1] M. R. Molas,[2] M. Zinkiewicz,[1] Ł. Bala,[1] K. Nogajewski,[2] M. Potemski,[2] A. Wysmołek,[1] and A. Babiński[1]

[1]*Faculty of Physics, University of Warsaw, ul. Pasteura 5, 02-093 Warsaw, Poland*

[2]*LNCMI CNRS-UGA-UPS-INSA-EMFL, 25 rue des Martyrs, 38042 Grenoble, France*

______________________________

[a)] Electronic mail: katarzyna.golasa@fuw.edu.pl

**Abstract**

Temperature-dependent (5 K to 300 K) Raman scattering study of $A_{1g}/A_1'$ phonon modes in monolayer (1L), bilayer (2L), trilayer (3L), and tetralayer (4L) MoTe$_2$ is reported. The temperature evolution of the modes' intensity critically depends on the flake thickness. In particular with λ=632.8 nm light excitation, a strongly non-monotonic dependence of the $A_{1g}$ mode intensity is observed in 2L MoTe$_2$. The intensity decreases with decreasing temperature down to 220 K and the $A_{1g}$ mode almost completely vanishes from the Stokes scattering spectrum in the temperature range between 160 K and 220 K. The peak recovers at lower temperatures and at *T*=5 K it becomes three times more intense that at room temperature. Similar non-monotonic intensity evolution is observed for the out-of-plane mode in 3L MoTe$_2$ in which tellurium atoms in all three layers vibrate in-phase. The intensity of the other out-of-plane Raman-active mode, (with vibrations of tellurium atoms in the central layer shifted by 180° with respect to the vibrations in outer layers), only weakly depends on temperature.

The observed quenching of the Raman scattering in 2L and 3L MoTe$_2$ is attributed to a destructive interference between the resonant and non-resonant contributions to the Raman scattering amplitude. The observed "antiresonance" is related to the electronic excitation at the M point of the Brillouin zone in few-layer MoTe$_2$.

______________________________

[a)] Electronic mail: katarzyna.golasa@fuw.edu.pl

# I. INTRODUCTION

Owing to layered structure of the crystal lattice, transition metal dichalcogenides (TMDs) possess a variety of remarkable electronic and optical properties [1-3]. Depending on the chemical composition, the TMDs can be metallic, superconducting or semiconducting. Upon reducing the thickness of semiconducting TMDs down to the monolayer limit, their band gap transforms from indirect to direct [4], which makes them promising candidates for potential applications in photovoltaics and optoelectronics [5-7]. Practical perspectives along with a widespread will to understand the physics of TMDs have recently turned their study into a very active and quickly developing area. One of the problems that remains to be elaborated is the role of electron-phonon interactions in these materials. In addition to being an interesting issue for fundamental research, the crystal lattice dynamics plays an important role for the performance of possible devices. Phonons are essential for several physical processes, *e.g.* carrier scattering [8], heat propagation [9], and mechanical strength of crystals [10]. A fascinating property of the lattice dynamics in few-layer TMDs is a critical dependence of the phonon energy spectrum on the material's thickness. This dependence concerns both the low-energy spectral range, in which the shear and breathing modes related to rigid layer displacements can be observed [11-12], as well as high-frequency modes, which involve vibrations of chalcogen and metal atoms. An example of the effect in the latter case is the evolution of the out-of-plane ($A_{1g}$) and in-plane ($E^1_{2g}$) modes with the layer thickness, which can be seen in Raman scattering measurements and, as recently shown, can be nicely described within a linear chain model [13]. The energy difference between these two modes is commonly used to estimate the thickness of a TMD film [14-15]. It is important to note that while the low-energy Raman-active modes can be well-modeled assuming rigid vibrations of all three monoatomic planes, the interactions between atoms in adjacent planes cannot be neglected in the description of high-frequency modes.

More information on the lattice vibrations and electron-phonon interactions can be provided by resonant Raman scattering measurements in which the resonance between the exciting photons and electronic excitations of the crystal is exploited [16-20]. A specific structure of the conduction and valence energy bands in semiconducting TMDs, which involve *p*-orbitals of chalcogen atoms and *d*-orbitals of metal atoms results in a complicated pattern of resonance effects. For example, the λ=632.8 nm light (1.96 eV), is known to enhance the Raman scattering related to the out-of-plane $A_{1g}$ mode in $MoS_2$ [16], [which is due to its resonance with the A and B excitons at low and room temperature respectively]. The resonant excitation also gives rise to strongly enriched Raman scattering spectra, which comprise several modes due to multiphonon processes mainly related to phonons at the edge of the Brillouin zone [21]. With the same excitation Davydov-split modes of the out-of-plane $A_{1g}$ vibrations can be observed in $MoTe_2$ [13, 22]. Moreover, an unexpected enhancement of the anti-Stokes to Stokes scattering intensity ratio of the spectral line related to the out-of-plane vibrations of Te atoms in thin layers of $MoTe_2$ can be seen [22].

In this report we study the evolution of the out-of-plane modes in monolayer (1L), bilayer (2L), trilayer (3L), and tetralayer (4L) $MoTe_2$ as a function of temperature from 300K down to 5 K. We show that the evolution critically depends on the flake thickness as well as on the characteristics of the involved vibrations. In particular, we observe a quenching of the $A_{1g}$ peak in the Stokes Raman scattering spectrum of 2L $MoTe_2$ excited with λ=632.8 nm light in the temperature range of 160 K – 220 K and a similar non-monotonic evolution of the higher-energy component of the A'$_1$ mode in 3L $MoTe_2$. We associate the effect with an antiresonance resulting from a destructive interference of the resonant and the non-resonant contribution to the Raman scattering amplitude.

## II. SAMPLES AND EXPERIMENTAL SETUP

The MoTe$_2$ samples under study were prepared on SiO$_2$ (320 nm)/Si substrates by means of polydimethylsiloxan-based mechanical exfoliation [23] of bulk crystals purchased from HQ Graphene. The thicknesses of obtained flakes were estimated with the aid of optical contrast measurements (see Fig. 1), intermittent-contact atomic force microscopy and Raman spectroscopy characterization.

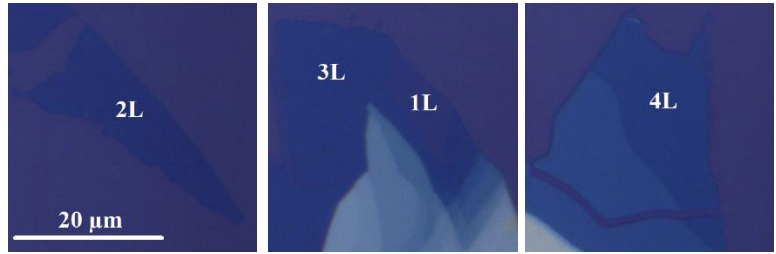

FIG. 1. Optical microscope image of 1L-4L MoTe$_2$.

The unpolarized Raman scattering measurements were carried out in the backscattering geometry with the use of three different continuous-wave laser light sources: (i) He-Ne ($\lambda$=632.8 nm, 1.96 eV), (ii) Ar-Kr ($\lambda$=647.1 nm, 1.91 eV), and (iii) Ar$^+$ ($\lambda$=514.5 nm, 2.41 eV). The Stokes and anti-Stokes (case (i) only) scattering spectra were recorded in a broad temperature range (5 K -300 K) on 1L, 2L, 3L, and 4L MoTe$_2$ flakes. The sample of interest was placed on a cold finger in a continuous flow cryostat, which was mounted on motorized x-y positioners. The excitation light was focused by means of a 50× long working distance objective giving the spot of around 1 μm$^2$ area. The laser power was kept at the level of 100 μW in order to avoid heating effects. The Raman signal was collected via the same microscope objective, sent through an 0.5 m or an 0.75 m monochromator and detected with a liquid-nitrogen cooled charge-coupled-device camera.

## III. EXPERIMENTAL RESULTS

The Raman scattering spectra observed in the 150 cm$^{-1}$ – 300 cm$^{-1}$ range for 1L, 2L, 3L and 4L MoTe$_2$, measured with λ=632.8 nm excitation are shown in Fig. 2. The peaks ascribed to the Raman-active out-of-plane $A_{1g}/A_1'$ and in-plane $E^1_g/E''$ modes, as well as the bulk-inactive mode $B_{1u}$, which becomes active in thin MoTe$_2$ layers ($A_{1g}/A_2'$) [24] can be seen in the Figure.

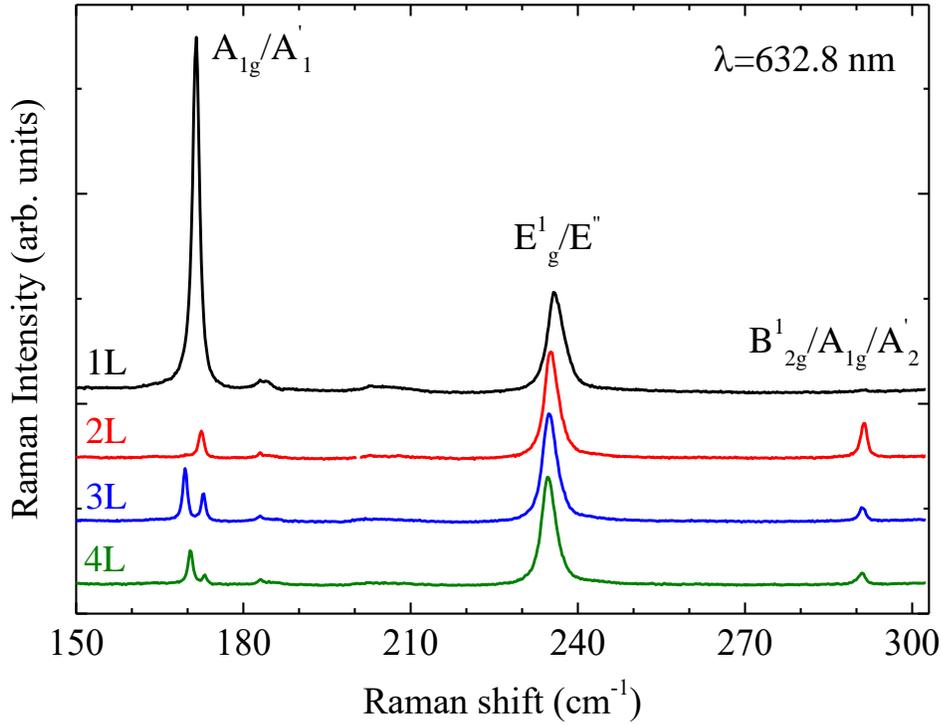

FIG. 2. Room-temperature Raman spectra of 1L, 2L, 3L and 4L MoTe$_2$ measured at room temperature with λ=632.8 nm excitation.

In what follows we focus on the out-of-plane vibrations, which give rise to the group of peaks at ~170 cm$^{-1}$. As one can notice in Fig. 2, the number of peaks forming that group depends on the number of possible Raman-active modes of the out-of-plane vibrations. The corresponding Raman-active vibrations are schematically presented in Fig. 3. There is one such mode of $A_1'$ ($A_{1g}$) symmetry in 1L (2L) MoTe$_2$ and two $A_1'$ ($A_{1g}$) modes in 3L (4L) MoTe$_2$ [13, 22, 25]. The number of spectral lines observed is related to the energy difference between Davydov-split components in NL MoTe$_2$. The splitting originates from van der Waals interactions between adjacent layers in the crystal lattice [13]. In the case of 1L MoTe$_2$, a single-molecule composition of the unit cell implies the existence of only one Raman-active mode of the out-of-plane vibrations. This mode (A'$_1$) gives rise to the appearance of an individual Lorentz-type line in the Raman scattering spectrum, as shown in Fig. 2. There are two modes in 2L MoTe$_2$ due to the out-of-plane oscillations of Te atoms in opposite directions with respect to the central Mo atom. The A$_{1g}$ mode is characterized by no phase difference between the vibrations of Te atoms in both planes - at a given moment they all move away from the central Mo atom in both planes. The other mode (A$_{2u}$), with 180° phase difference between the oscillations of Te atoms in the planes surrounding the Mo plane is infrared-active. As a consequence only one peak due to the out-of-plane vibrations should be expected in the Raman scattering spectrum of 2L MoTe$_2$, which is indeed the case as shown in Fig. 2. There are two Raman active modes in 3L MoTe$_2$ (see Fig. 3). Te atoms in all three layers of the crystal oscillate in-phase in the first mode, which will be referred to as the (*i*) mode. In the other Raman-active mode, the vibrations in the central layer are phase-shifted by 180° with respect to the vibrations in the outer layers. The mode will be referred to as the (*j*) mode. Similarly in 4L MoTe$_2$, the (*i*) mode is related to in-phase vibrations in all layers while the (*j*) mode comprises vibrations in the central two layers shifted by 180° as compared to vibrations in the outer layers. As it was previously shown, the number of the out-of-plane vibrations increases with the flake's thickness and the evolution can be understood in terms of interactions between the layers of TMD crystals [13]. Ultimately, in bulk MoTe$_2$, two out-of-plane modes are expected which are similar to the modes in 2L MoTe$_2$. The mode with Te atoms vibrating in-phase in both planes (A$_{1g}$) is Raman-active. The other mode, with the 180° phase difference between the oscillations of Te atoms in adjacent planes ($B_{1u}$) is inactive.

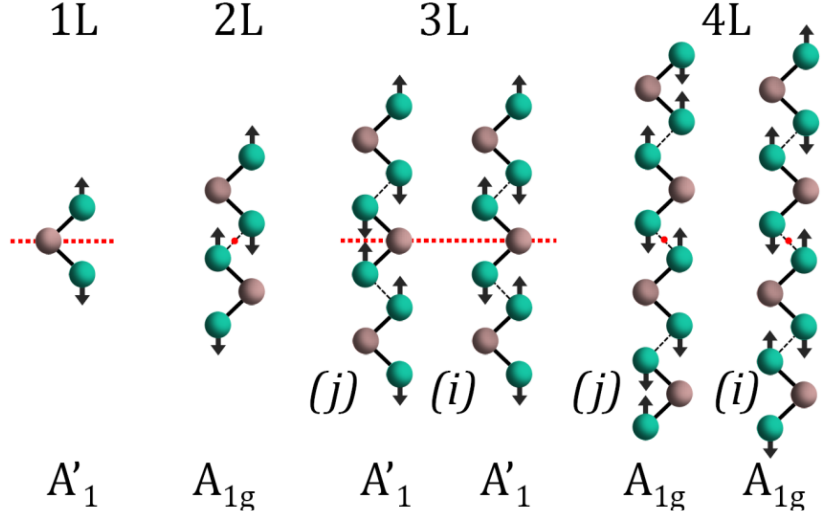

FIG. 3. Schematic representation of the Raman-active out-of-plane vibrational modes in 1L, 2L, 3L, and 4L MoTe$_2$. Dashed lines denote the mirror symmetry plane and red points mark the inversion centers. The $A_1'$ and $A_{1g}$ correspond to odd and even number of layers of the crystal. The modes with Te atoms vibrating in-phase in all layers for 3L and 4L MoTe$_2$ are denoted with (*i*). The other Raman-active modes in those structures are denoted with (*j*).

The temperature evolution of the Raman scattering spectrum due to out-of-plane modes in 1L, 2L, 3L and 4L MoTe$_2$ excited with $\lambda$=632.8 nm ($\lambda$=647.1 nm ) is shown in the upper (lower) panel of Fig. 4.

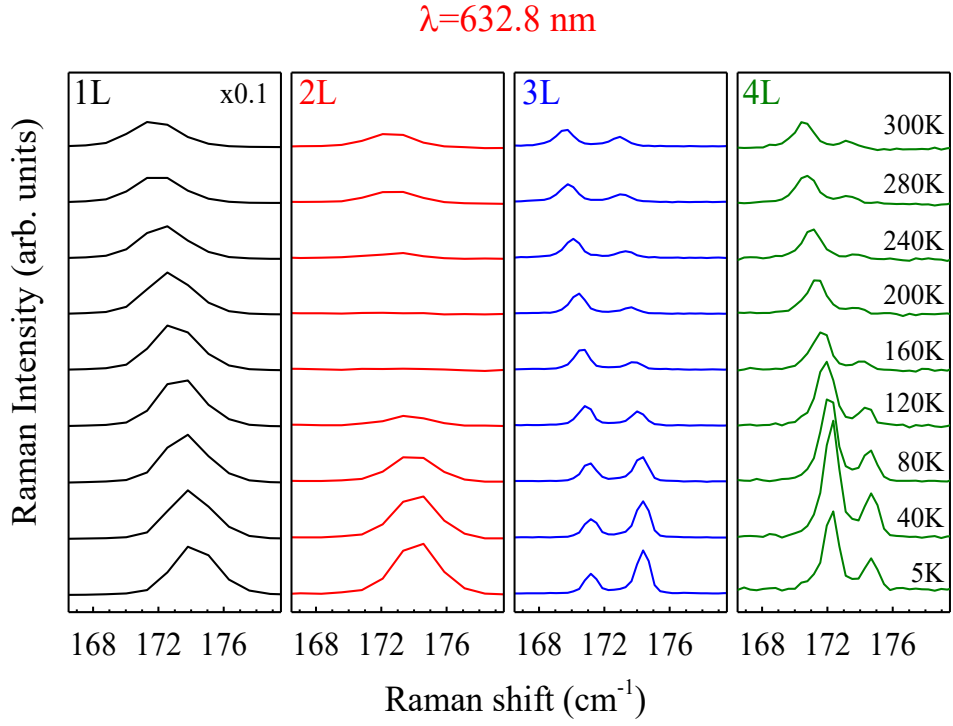
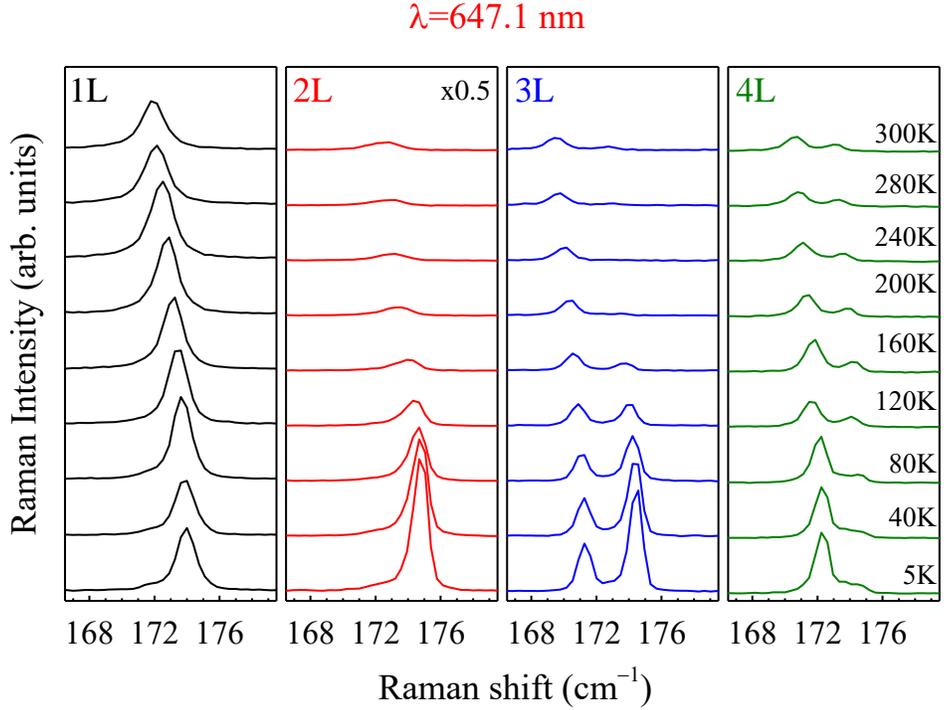

FIG. 4. Temperature dependent Raman spectra of 1L, 2L, 3L and 4L MoTe$_2$, excited with 1.96 eV ($\lambda$=632.8 nm) – upper panel and 1.91 eV ($\lambda$=647.1 nm) – lower panel.

As can be seen in Fig. 4, the evolution strongly depends on the number of layers, which form the structure. In order to capture main features of the evolution, the relative [$(I_{A_{1g}}/I_{E_g^1})$ for even number of layers or $(I_{A_1'}/I_{E''})$ for odd number of layers] intensities of the $A_{1g}/A_1$' – related peaks in 1L, 2L, 3L, and 4L MoTe$_2$ in the Raman scattering are summarized in Fig. 5. The relative scale allows to focus on the most important properties of the evolution as the $E^1{}_g/E^{''}$ in-plane vibrational mode does not exhibit any resonant behavior and only weakly depends on temperature.

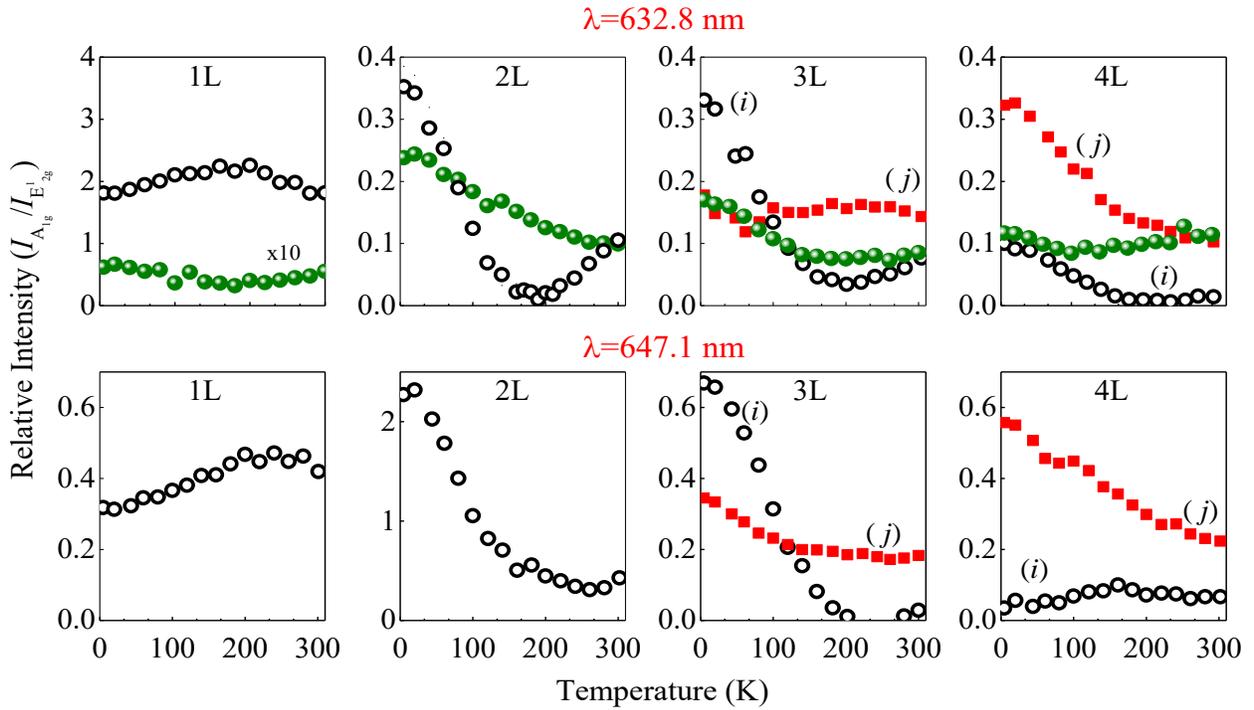

FIG. 5. Temperature dependence of the relative intensities of the $A_{1g}/A_1$' – related peaks in 1L, 2L, 3L, and 4L MoTe$_2$ in the Raman scattering spectra excited with 1.96 eV ($\lambda$=632.8 nm) – upper panel and 1.91 eV ($\lambda$=647.1 nm) – lower panel. Data for the mode with in-phase vibrations are shown with open black circles, data for the (*j*) mode in 3L and 4L MoTe$_2$ are shown with closed red squares. For the sake of comparison, the temperature dependence of the relative intensities of peaks due to in-phase $A_{1g}/A_1$' modes in the Raman scattering spectra excited with 2.41 eV ($\lambda$=514.5 nm) are also shown with closed green circles.

A variety of temperature evolution of the $A_{1g}/A_1$' can be appreciated in Fig. 4. Let us focus first on the Raman spectra obtained under $\lambda$=632.8 nm excitation. There is just one out-of-plane mode in 1L MoTe$_2$. The relative intensity of the A$_1'$ peak observed in 1L MoTe$_2$ only weakly depends on temperature with a maximum at $T$=200 K. The most striking feature of the results presented in Fig. 5 is the apparent annihilation of the A$_{1g}$ peak from 2L MoTe$_2$ in the temperature range 160 K–220 K. In consequence, three distinct temperature ranges of the temperature dependence can be distinguished: (i) the low-temperature enhancement of the A$_{1g}$ peak, (ii) its annihilation at 160 K–220 K, and (iii) its recovery towards room temperature. Similar, although less pronounced minimum in the A$_1'$ (*i*) peak intensity can be observed around 200 K in 3L MoTe$_2$. Moreover, a distinct evolution pattern can be found for the (*i*) and (*j*) modes in 3L MoTe$_2$. The relative intensity of the out-of-plane modes in 3L MoTe$_2$ also changes with temperature: while at room temperature the low-energy (*j*) peak is stronger than the high-energy (*i*) peak, the latter peak dominates the spectrum at $T$=5 K. In 4L MoTe$_2$ both (*i*) and (*j*) modes gain in intensity with decreasing temperature.

Results of measurements performed under $\lambda$=647.1 nm excitation follow in general those of measurements at $\lambda$=632.8 nm with three points to be noted. First, the observed features of the dependence (the maximum for 1L and minima in 2L and 3L) shift to higher temperature with the change in excitation from 1.96 eV to 1.91 eV. This may reflect the temperature evolution of the MoTe$_2$ band gap, which shrinks with increasing temperature. Next, there is a substantial difference between the intensity of peaks at $\lambda$=647.1 nm as compared to those obtained at $\lambda$=632.8 nm. While for 1L MoTe$_2$, the former excitation leads to much lower intensity than in the latter case, the opposite effect can be observed for 2L - 4L MoTe$_2$. Finally in 4L MoTe$_2$ only the (*j*) mode gains intensity with decreasing temperature. The intensity of the other (*i*) out-of-phase mode only weakly depends on temperature.

## IV. DISCUSSION

The amplitude of the non-resonant first order Raman scattering process can be modelled within the polarizability model. It is related to the change of the polarizability of atomic bonds under atomic displacement. The expected Raman spectra obtained within the polarizability model were calculated for few-layer MoTe$_2$ in Supplementary Information to Ref. [13]. As far as the out-of-plane A$_{1g}$/A$_1$' vibrations are concerned, the spectra are dominated by the in-phase (*i*) mode. Other [ (*j* ),...] modes are hardly seen in the theoretical Raman scattering spectra. The calculated spectra correspond well to experimental results obtained under λ = 514.5 nm excitation, which is far from resonance with any maxima in joint density of states in few-layer MoTe$_2$ [13, 21]. Moreover, at λ = 514.5 nm (2.41 eV) excitation (see green points in Fig. 5), the relative intensity of the in-phase A$_{1g}$/A$_1$' modes does not significantly change upon the modulation of temperature down to *T*=5 K. This confirms the non-resonant character of Raman scattering with that excitation in a broad temperature range.

This description within the polarizability model must be expanded when the Raman scattering excitation approaches the energy corresponding to Van Hove singularity in semiconductor energy structure. The Raman scattering amplitude in such a case depends on both resonant and non-resonant contributions [26]:

$$P_{ph}(\omega_i) \sim \left| \frac{\langle 0|H_{eR}(\omega_i)|a\rangle\langle a|H_{e-ph}|a\rangle\langle a|H_{eR}(\omega_s)|0\rangle}{(E_a - \hbar\omega_i)(E_a - \hbar\omega_s)} + C \right|^2 \quad (1)$$

where $H_{eR}$ ($H_{e-ph}$) is the electron-radiation (electron-phonon) interaction Hamiltonian, $|0>$ is the ground state of a semiconductor with no excited electron-hole pairs, $|a>$ is the resonant intermediate state with energy $E_a$, $\hbar\omega_i$ ( $\hbar\omega_s$ ) denotes the energy of incident laser (scattered) light, and C is a constant non-resonant background. The non-resonant term corresponds to the polarizability of bonds, resulting from rearrangement of virtual electronic states. The resonant term corresponds to the polarizability of bonds, which corresponds to electronic transition from the valence band to the conduction band affects the polarization. Those two processes have different origins and the second term in Eq (1) does not correspond to the first term far from the resonance. Because of the denominator, the whole first term disappears in such a condition and it does not evolve into the non-resonant term. Only the second term persists. As it is shown in Ref. [26] both terms must be added before the sum is squared, which is at the basis of our model.

The resonant enhancement of Raman scattering due to the resonance of the incoming laser or scattered light energy with the energy of electronic state was observed in several TMDs [16-20, 27]. A key factor for the resonant behavior is the electron-phonon coupling, which is described by the corresponding $H_{\text{e-ph}}$ matrix element in the first term in Eq. (1). The different coupling of the $A_{1g}/A_1$' and $E^1_g/E$'' phonons with electronic transition is usually explained in terms of different directions of atomic displacements in the unit cell, which correspond to the vibrations. For example in $MoS_2$ the character of the final state of the direct electronic transition resulted from red light excitation is associated with the $d_{z2}$ orbitals of the Mo ion, which point perpendicular to the atomic planes. Atomic vibrations in that direction as in the $A_{1g}/A_1$' modes strongly affect the polarizability while displacements of atoms in the parallel direction as in the $E^1_g/E$'' mode has only a minor effect. As a result, the laser excitation in resonance with the A or B excitons in semiconductor TMDs is known to enhance the scattering on the out-of-plane $A_{1g}$ modes and not on the in-plane $E^1_{2g}$ modes. [16]

Under resonance conditions, the contribution of the non-resonant terms C to the scattering probability can be regarded as constant. The two (the non-resonant and the resonant) components in Eq. (1) are added before their sum is squared to obtain the scattering probability. Therfore a destructive interference of those components can take place if the relative signs of those two components are opposite. A resonant quenching of the Raman scattering instead of its enhancement can occur in such a case. The effect was observed in several semiconductor systems, such as wurtzite ZnS [28], CdS [29, 30] or pyrite $RuS_2$ [31], however up to our knowledge it was not reported previously in TMDs. The antiresonance is responsible in our opinion for the annihilation of the Raman scattering at λ=632.8 nm due to the $A_{1g}$ mode in 2L $MoTe_2$ and for the quenching of the in-phase (*i*) mode of the $A'_1$ phonon in 3L $MoTe_2$.

A strict theoretical analysis of the observed effect demands a more detailed approach which is beyond the scope of our experimental work. There are, however, some points which can be addressed based on previous studies of lattice dynamics in $MoTe_2$. First is the identification of the Van Hove singularity in the $MoTe_2$ energy structure, which may be responsible for the resonance. In our opinion, this is related to the maximum of the electronic density of states in the highest valence band and the second lowest conduction band at the M point of the Brillouin zone [32]. The maximum coincides with the energy of the excitation with λ=632.8 nm (1.96 eV). The excitation was shown to result in several additional features in Raman scattering spectrum of few-layer $MoTe_2$ due to a double resonance Raman scattering. The features were also clearly observed in the Raman spectrum of 2L $MoTe_2$ measured at *T*=5 K (see Fig. 6). They were proposed to be related to the double resonance Raman modes: $\omega_3$ [$E^2_g$ (M)+TA(M) or 2LA(M)], $\omega_4$ [$A_{1g}$(M)+TA(M)], $\omega_5$ [2$A_{1g}$(M) or $E^1_g$ (M)+TA(M)], and $\omega_6$ ($E^1_g$ (M)+LA(M)). [32]

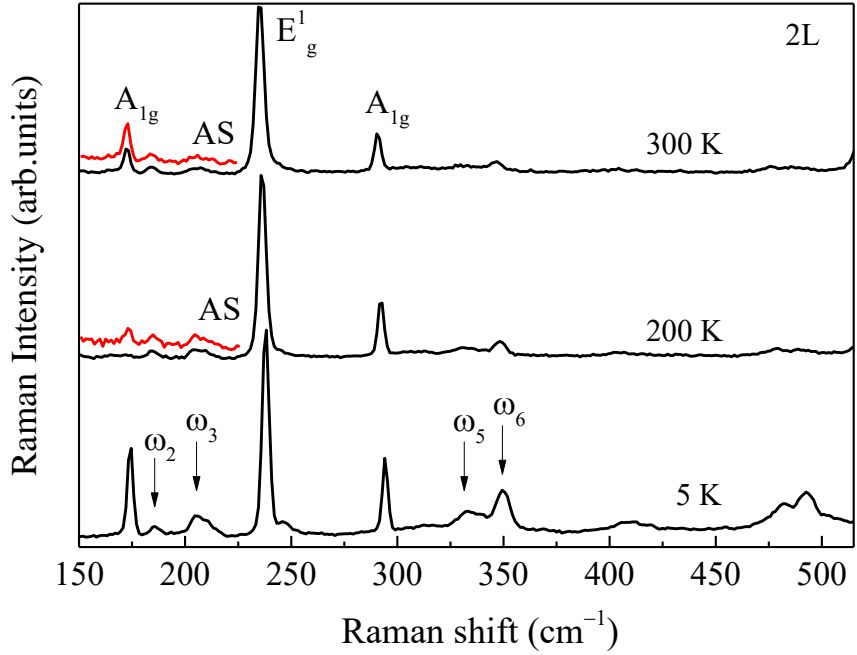

FIG. 6. The Stokes Raman scattering spectra of 2L excited with 1.96 eV ($\lambda$=632.8 nm) laser light at $T$=5 K, 200 K, and 300 K. The anti-Stokes scattering spectra at 200 K and 300 K, denoted with AS, are also presented.

Next is the question why only the in-phase modes (in 2L and 3L) are sensitive to the antiresonance effect. The antiresonance may be expected when the contributions to deformation potential due to the non-resonant and resonant (involving real electronic excitation) components of the Raman scattering amplitude have opposite signs. In particular, the different effect of temperature on the (*i*) mode and the (*j*) mode in 3L, with neither antiresonance nor resonance observed in the latter case may suggest quite different contribution of the two vibrational modes to the deformation potential. A stronger effect on the (*i*) mode suggests its stronger coupling with the electronic excitation. We relate the difference to the composition of the electronic wave functions at the M point of the MoTe$_2$ Brillouin zone, which are proposed to be involved in the resonance. As it was shown in Refs [34, 35], the valence band edge states at the M point in MX$_2$ material have substantial component of the $p_z$ X (chalcogen) atom orbitals. Moreover, the second lowest minimum of the conduction band at the M point in $MX_2$ has substantial component of the $d_z^2$ M (metal) atom orbitals. Both orbitals point along the *c*-axis of the crystal, which should lead to the coupling of the corresponding electronic excitation to the out-of-plane $A'_1/A_{1g}$ vibrational modes. Our experimental results show, however, that crucial for the antiresonance is the relative phase of vibrations in layers constituting the structure. In the (*i*) mode the distance between Te atoms in adjacent layers modulates and affects the interaction between Te-related $p_z$ orbitals. This may lead to substantial modulation of the $MoTe_2$ band structure. In the (*j*) mode of the out-of-plane vibrations Te atoms in the central (two central) layers vibrate out-of-phase with respect to the outer layers in 3L $MoTe_2$. The modulation of the distance between Te atoms in adjacent layers is weaker, which may explain the weaker electron–phonon coupling.

The resonant process could also explain the difference between the results obtained at λ=632.8 nm and λ=647.1 nm excitation. Temperature of the minima in the $A_{1g}$/A' intensity for 2L (3L) $MoTe_2$ shifts by approximately 80 K (40 K) with the change in the excitation energy from 1.96 eV (λ=632.8 nm) to 1.91 eV (λ=647.1 nm). The change may correspond to the expected band gap energy change due to modulation of temperature, although the estimated temperature coefficients (1.6 meV/K for 2L and 0.8 meV/K for 3L) are rather high as compared to the observed linear temperature coefficient of the A excitons in $MoTe_2$ [36].

Finally, the interference of the resonant and non-resonant processes may also explain the enhancement of the anti-Stokes to Stokes scattering intensity ratio of the spectral line attributed to the out-of-plane vibrations of Te atoms in thin layers of $MoTe_2$. The weakening of the $A_1'$ (*i*) mode (due to in-phase vibrations) in the Stokes Raman scattering in 3L $MoTe_2$ at room temperature has been recently reported [22]. As it can be seen in Fig. 6, the effect is even more pronounced in 2L at T ≈ 200 K. While the Stokes $A_{1g}$ mode is quenched, the $A_{1g}$ mode can be well distinguished in the anti-Stokes Raman scattering at λ=632.8 nm excitation. This suggests that the effect previously reported in 3L MoTe$_2$ [22] at room temperature results from the resonant quenching of the A$_1$' mode at lower temperature.

## CONCLUSIONS

We have studied the temperature-dependent Raman scattering spectra of 1L, 2L, 3L, and 4L MoTe$_2$ at temperatures between 5 K and 300 K. We found that the temperature evolution of the modes' intensity critically depended on the flake thickness. In particular with λ=632.8 nm light excitation the Stokes Raman scattering intensity due to the out-of-plane A$_{1g}$/A'$_1$ in-phase vibrations exhibits a non-monotonic temperature dependence in 2L and 3L MoTe$_2$. The quenching of the in-phase A$_{1g}$/A'$_1$ modes is related to an antiresonance resulting from the cancellation of the resonant and non-resonant contributions to the Raman scattering amplitude. The resonance of the excitation light with the electronic transition at the M point of MoTe$_2$ Brillouin zone is proposed to account for the observed effect.

## ACKNOWLEDGMENTS

This work has been supported by the National Science Centre under grants no. DEC-2013/11/N/ST3/04067, DEC-2013/10/M/ST3/00791 and DEC-2015/16/T/ST3/00496. Funding from European Graphene Flagship and European Research Council (ERC-2012-AdG-320590-MOMB) is also acknowledged.